# Low field magnetoelectric effect in Fe substituted $Co_4Nb_2O_9$

C. Dhanasekhar[1,3], S. K. Mishra[2], R. Rawat[2], A. K. Das[1] and A. Venimadhav[3*]

[1]*Department of physics, Indian Institute of Technology, Kharagpur -721302, India*

[2]*UGC-DAE Consortium for Scientific Research, Khandwa Road, Indore, 452017, Madhya Pradesh, India*

[3]*Cryogenic Engineering Centre, Indian Institute of Technology, Kharagpur -721302, India*

## Abstract

$Co_4Nb_2O_9$ (CNO) having α-$Al_2O_3$ crystal structure with Co chains along c-direction shows gigantic magnetoelelctric coupling below antiferromagnetic ordering temperature of 27 K but above a spin flop field of 1.6 T. We have investigated structural, magnetic and magnetoelectric properties of Fe substituted (10% and 20%) samples and compared with the parent one. In fact magnetic and specific heat measurements have revealed an additional magnetic transition below 10 K and presence of short range magnetic ordering above ~ 50 K in parent as well as in Fe substituted samples. Linear magnetoelelctric and ferroelectric behaviours are evidenced in the Fe substituted samples where an electric field of 5 kV/m is sufficient to align the dipoles and the magnetoelelctric coupling is ensured for magnetic fields as low as 0.25 T, far below the spin flop field.

\* Corresponding author, venimadhav@hijli.iitkgp.ernet.in

**INTRODUCTION**

Manipulation of magnetization for data storage by electric field is the immediate interest in spintronic devices. Materials with large magnetoelectric coupling are desired for such applications. Magnetoelectric (ME) effect has been found in two kinds of materials: the well known multiferroic materials and linear magnetoelectric materials like $Cr_2O_3$, $NdCrTiO_5$, $MnTiO_3$, etc [1-3]. The major difference between these two classes is that the later ones have no spontaneous polarization, but display linear variation in electric polarization in varying magnetic field. In the case of multiferroics, type –II systems are attractive as the electric order parameter is driven by the magnetic ordering [4-10].

The family of $A_4M_2O_9$ (A= Ta, Nb; M= Co, Mn, Fe) are attractive candidates [11-15] for ME applications. The crystal structure of (CNO) is similar to the α-$Al_2O_3$ with space group of $P\bar{3}c1$ and the unit cell is built up by the three-unit cell of corundum structure. In CNO, the Co and Nb-ions occupies the Al sites with 2:1 ratio, and the Co-ions have two different crystallographic sites. Though not much of ME studies have been carried out on these systems, ME properties of $Co_4Nb_2O_9$ (CNO) have been attempted by few groups. In powder samples of CNO, Kolodiazhnyi *et al.* have observed magnetodeielctric effect in the vicinity of magnetic ordering temperature and the dielectric constant increases above a certain critical magnetic field (> 12 kOe) and this magnetic field is assigned to the spin flop field [15]. Later, Fang *et al.* reported a large ME coupling (ME=18.4 ps/m) in CNO powder samples and interestingly, the ME coupling appears above the spin flop field of 15.7 K Oe [16]. Ren *et al* have synthesized single crystals of CNO and studied the magnetic properties along a- and c-axes, while these authors have observed a spin flop field of 7.5 K Oe along a-axis [17].

The magnetic structure and spin orientation of the CNO is still unsettled**.** Recently Khanh *et al.* studied the magnetic structure and ME coupling in CNO single crystal [18]. The authors have proposed that the magnetic moments lie along the trigonal basal plane and assigned the magnetic symmetry to C2/*c'*, which is different from the early proposed magnetic symmetry of $P\bar{3}\ c'1$ [19] where $Co^{2+}$ magnetic moments were considered aligned along the trigonal axis(c-axis). In the single crystal, the spin flop was found along the [$1\bar{1}0$] direction and the direction dependent polarization measurements with linear ME (30 ps/m). Compared to reported bulk samples, single crystals have shown lower spin flop field and larger ME coupling. In the present paper, we have partially substituted Co by Fe in CNO and found a marginal increase in antiferromagnetic ordering temperature; more significantly, a linear ME coupling at low magnetic fields much less than the spin flop field is noticed.

**Experimental**

Polycrystalline CNO and $Co_{4-x}Fe_xNb_2O_9$ (CFNO) (x= 0.1 and 0.2) samples were prepared by the standard solid-state reaction method using $Co_3O_4$, $Fe_2O_3$, and $Nb_2O_5$. Stoichiometric mixtures of these powders were grinded thoroughly, heated in air at 900 $^oC$ and 1100 $^0C$ for 12 h, finally pressed hydrostatically into pellets and sintered at 1100 $^oC$ for 12 h. The crystal structures of the samples were characterized by powder X-ray diffraction (XRD) and analysed using the full proof program. Measurements of the magnetization were conducted using a Superconducting quantum interference device magnetometer (SQUID of Quantum Design, USA). Semi adiabatic heat pulse technique is used to measure the specific heat in temperature range of 2–100 K and in 0 and 5 T magnetic fields. Temperature dependent polarization measurements were performed using Keithley 6517A electrometer on samples of thickness 0.5 mm in capacitor geometry using silver paste contacts on both the sides. In details, the samples were first cooled to 70 K, which is above the ordering temperature for all samples. First, sample is annealed from 70 K to 10 K in presence of both electric and magnetic fields and at 10 K the electric fields is switch off and short-circuited the electrodes for long-enough time to avoid the residual surface charge effects. Sample is than warmed with 5 K/min and recorded the pyroelectric current in the presence of magnetic field. Finally, the temperature dependent polarization is obtained by integrating the pyrocurrent with respect to the time.

**RESULTS AND DISCUSSION**

**Crystal structure at room temperature**

The crystal structure of the CNO sample is shown in the figure.1(a), where the two different Co sites (Co1and Co2) are shown with the open and closed distorted octahedra made of oxygen atoms. The Co (1) and Co (2) octahedra face shared and Co (2) octahedra are edge shared. The Co (1) octahedra are corner shared by oxygen atoms. The powder XRD pattern of the CNO and CFNO were investigated by Rietveld analysis and the structure of all the samples have been assigned to trigonal crystal structure with space group of $P\bar{3}c1$. Typical XRD patterns of x = 0 and x = 0.20 at room temperature are shown in the figure.1 (b, c) and they are free from parasitic phases. The difference between the experimental and calculated diffraction patterns is shown in the bottom of the figure with blue lines and the corresponding Bragg peaks are shown with vertical pink lines. The corresponding lattice parameters for the x = 0 and x = 0.20 samples is shows in the inset of figure 1(b, c). With Fe substitution one can infer that the c lattice parameter increases while a- lattice parameter contracts by small amount such that the overall volume reduces which is consistent with the smaller size of Fe.

**Magnetic characterization**

Temperature dependent field cooled magnetization, measured under H = 500 Oe for the pure sample is shown in the inset of figure 2 (a). The behavior matches well with the previous reports. However, the field cooled magnetization measured at 50 Oe on the pure sample shows a sharp rise at 50 K, followed by the antiferromagnetic (AFM) transition at 27 K and further below 10 K, an upturn in magnetization is found. The transitions at 50 K and 10 K were not discussed before. The inset of figure 2 (a) shows that the transition at 50 K gets suppressed for higher magnetic fields. The inverse susceptibility data at 50 Oe is fitted with the Curie-Weiss (CW) law above the 50 K, which is shown in the inset of figure 2 (b). The fitting gives a Curie-Weiss temperature ($\theta_{CM}$) of = -113.8 K and an effective moment of 9.45 $\mu_B$ /f.u. The obtained moment is higher than the spin only moment of $Co^{2+}$ (S = 3/2 and $\mu_{cal}$ = $[4\mu^2_{Co2+}]^{1/2}$ = 7.74 $\mu_B$) that suggests the orbital moment is not quenched in the parent sample. The observed values are in agreement with the previous reports [15, 18].

In CFNO samples, we have observed a ferromagnetic like behavior in the FC magnetization measurements for 50 Oe and the corresponding M vs. T plots are shown in the figure .2 (b) and (d). The antiferromagnetic ordering is slightly shifted to higher temperature ~32 K with Fe substitution and another magnetic transition is observed at 58 K in both samples. The transition at 10 K is also observed with Fe substitution and for clarity, is shown in the inset of figure 2 (c) for x = 0.1 sample. However, both 10 K and 58 K transitions get suppressed at higher fields in both samples and is depicted for x = 0.20 case in inset of figure 2 (d). The Curie-Weiss fit to Fe substituted samples are also shown in figure 2 (b), the C-W temperature ($\theta_{CM}$) is -131.2 K for x = 0.10 and -141.5 K for x = 0.2; and the effective moment is 9.36 $\mu_B$ and 10.69 $\mu_B$ for x = 0.1 and 0.2 samples respectively. These values do not match with the calculated spin only moments per formula unit, $\mu_{cal}$ = 7.79 $\mu_B$ and 7.85 $\mu_B$ for x = 0.1 and x = 0.2 samples [18]. The obtained parameters indicate that the dominant antiferromagnetic nature in both Fe substituted samples and unquenched nature of orbital moments for the both $Co^{2+}$ and $Fe^{2+}$ ions.

The M-H loop of the Fe substituted samples measured at 5 K and 30 K is shown in the figure 3 (a). A clear magnetic hysteresis is observed at low temperature, but with increasing temperature the width of the hysteresis decreases and disappears above 35 K. We further confirm the spin flop transition of the Fe substituted samples by measuring the magnetization under magnetic field at 5 K and 20 K and is shown in the figure 3 (b, c). The critical spin flop field is noticed at 1.6 T for Fe substituted samples and is shown with a vertical line and this value is indeed similar to that of parent sample. The observed ferromagnetic behavior, can be related to the uncompensated magnetic interactions between the intra and inter $Co^{2+}$ chains [20, 21] or canting of Co spins away from the basal plane [18] Further, the calculated frustration order parameter (f = $\theta_{CM}/T_N$) for CNO and CFNO samples is close to 5. This signifies the competing nature of magnetic interactions between the two Co sites.

**Specific heat measurements**

The temperature and magnetic field dependent specific heat data is shown in the figure 4 (a. b). Under zero magnetic field (shown with black line) Cp/T vs. T shows a λ- type peak at the antiferromagnetic transition (27 K), which indicates the long range order in the system. Interestingly, the peak width and amplitude does not vary under magnetic field of 5 T that suggests a strong antiferromagnetic nature. In Cp/T vs. T curve shows a clear shoulder at 10 K and no anomaly is found around 50 K. Similar to the parent sample, Fe substituted samples also show λ- type peak but much broader than parent sample in the vicinity of magnetic ordering temperature ~ 32 K. In the substituted samples as well, a shoulder is observed at 10 K indicating the possible long range ordering at low temperatures. Interestingly, in either of the samples, no anomaly is observed above 50 K; however, an increase in magnetization is noticed in parent sample and this enhancement pronounced with Fe substitution signifies the possible short range ferromagnetic order at higher temperatures.

**Pyroelectric measurements**

Figure 5 (a) shows the electric polarization of 20 % Fe substituted sample. It can be noticed that with increasing external magnetic field, the polarization increases linearly and reaches a maximum value of 31 µC/m$^2$ under 5 T. Like that of parent sample, ferroelectric behavior is coupled with the 32 K AFM transition. Though the saturation polarization is small in CFNO when compared to the parent one , a substantial pyroelectric current has been found at magnetic field as low as 0.25 T and is shown in the inset of figure 5 (a). The sign of the pyroelectric current can be reversed by reversing the poling field in the Fe substituted sample and is shown in the figure 5 (b). Further, electric polarization for different poling fields under a constant magnetic field of 5 T at 10 K for x = 0.20 samples is shown in the figure 5 (c). It is clear from figure 5 (c) that even a small poling field of 5 kV/m is sufficient to polarize the electric dipoles. The electric polarization saturates for poling field of 200 kV/m and does not change for a maximum applied poling field of 450 kV/m. This behavior is similar to the recently reported single crystal CNO, where the electric polarization gets saturated above 50 kV/m in (1 1 0) and (1 $\bar{1}$ 0) directions [17]. Figure 5 (d) shows the comparisons of the electric polarization under different magnetic fields for CNO and CFNO samples at 10 K. In contrast to the CNO sample, CFNO samples shows the significant polarization at lower fields in spite of having similar spin flop field that of pure sample. The ME coupling of the pure sample is found to be ~19.3 ps/m at lower fields (between 1-3 T) and decreases for higher fields (> 3 T). While in Fe substituted samples, ME is linear and has observed maximum ME coupling of 7.6 ps /m for 20% Fe substitution. The ME susceptibility of our powder samples is larger than the well know ME materials like $Cr_2O_3$ (4 ps/m), $MnTiO_3$ (2.6 ps/m), $NdCrTiO_5$ (0.51 ps/m) [1-3].

The results demonstrate that with Fe substitution ME coupling can be obtained at lower magnetic fields without significant change in the crystal structure. Furher, an additional magnetic ordering below 10 K and possible ferromagnetic (FM) short range correlation above AFM transition up to ~50 K is found in the Fe substituted samples. Though the saturation polarization is found decrease in Fe substituted samples, observation of low field ME coupling independent of spin flop is more promising. The microscopic origin of the linear ME coupling in powder CNO samples is explained by the domain effect [22] and in single crystal study Khanh *et al* have suggested that the liner ME can originate from the response of orbital magnetic moments as proposed by Scaramucci *et al* [23]. On the other hand from, strain measurements by Xie *et al* [24] showed a significant magneto elastic coupling in powder CNO sample below $T_N$ that signifies the magnetostriction mechanism as a possible reason for ME coupling. In parent sample the sharp increase of the electric polarization above 1-2 T is a signature of torodial moments [22]. The torodial moments exist due to the antisymetric terms (off diagonal terms) in the magneto electric tensor above the spin flop field. This behavior resembles with the toroidal moments in spin flop phase of $Cr_2O_3$ [22-23]. However, in $Cr_2O_3$, ME has been observed even for the fields less than the spin flop field. In Fe substituted sample, the increase in electric polarization with magnetic field is not as sharp as that of the parent one, this indicates the dilution or suppression of torodial moments. It suggests that the effect of disorder by Fe locally weakens the AFM coupling. Though at present the origin of low field ME effect is unclear, local canting of the moments due to disorder seems to play a role at low fields through Dzyaloshinskii-Moriya interaction. This study signifies a method of obtaining low field ME in CNO with substitution at Co sites.

In summary we have investigeted the structural, magnetic and ME properties of CNO, 10% and 20% Fe substituted samples.The magnetic study unvils additional magnetic transitions at 10 K and above 50 K, and specific heat measurement has suggested the possibility of long range ordering below 10 K. Short range ferromagnetic corrolations seems responsible for understanding the magnetic hysterisis behavior. Though Fe substitution has no effect on spin flop field, ME response is found in low magnetic field and electric fields which is promising for spintronic applications.

**ACKNOWLEDGMENTS :** The authors acknowledge DST- FIST facility in Cryogenic Engineering Centre and IIT Kharagpur funded VSM SQUID magnetometer. Dhanasekhar thanks MHRD, Delhi for senior research fellowship (SRF).

**Figure captions**

Figure 1. (a) Shows the crystal structure of CNO. Room temperature XRD and Rietveld refinement of (b) CNO and (c) CFNO (x=0.20); inset shows the lattice parameters.

Figure 2. (a) Shows the temperature dependent field cooled warming magnetization measured under 50 Oe for CNO sample and the inset shows the magnetization measured under 500 Oe; (b) Curie-Weiss fit for CNO and CFNO samples, (c) and (d) shows the M vs. T for CFNO samples. The inset of (c) shows first derivative of the magnetization for x= 0.10 and inset of (d) shows the magnetization vs. temperature of x = 0.20 samples at different magnetic fields.

Figure 3. (a) M-H of CFNO samples at 5 K; inset shows the M-H at 30 K. (b) and (c) shows the initial curves of CFNO samples measured at different temperatures and inset of (b) show the initial curves at different temperatures of parent CNO sample.

Figure 4. The Cp/T vs. T of (a) CNO (b) CFNO (x=0.20) samples as a function of temperature under 0 T and 5 T magnetic fields; magnified view at low temperature for x 0.20 is shown in the inset of (b).

Figure 5. For x = 0.20 sample, (a) shows temperature and magnetic field dependent polarization; inset shows the polarization at a low magnetic field, (b) shows switching of electric polarization and (c) displays electric polarization at different poling fields under 5 T. The comparison of the electric polarization for CNO and CFNO samples at 10 K is given in (d).

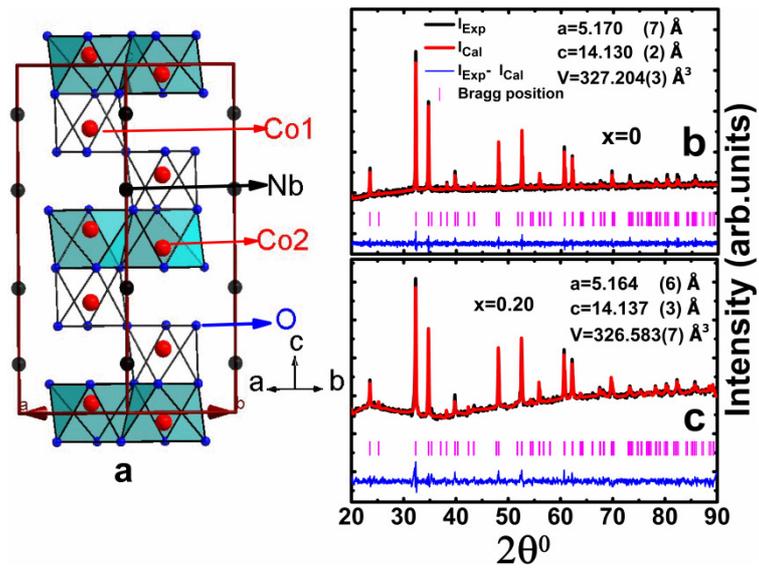

**Figure. 1**

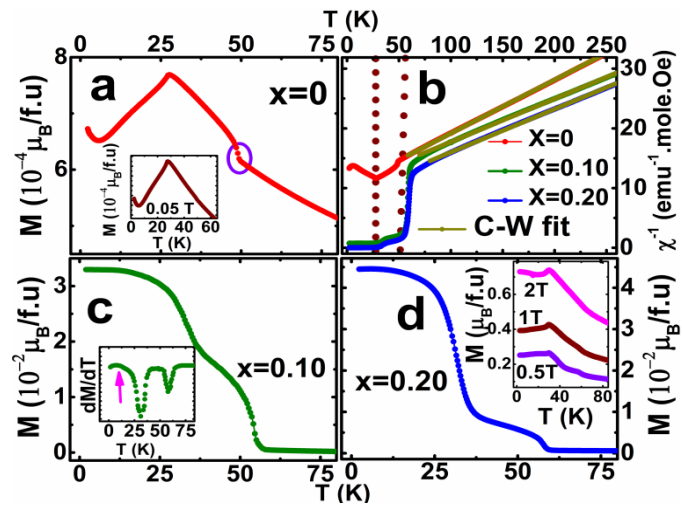

**Figure. 2**

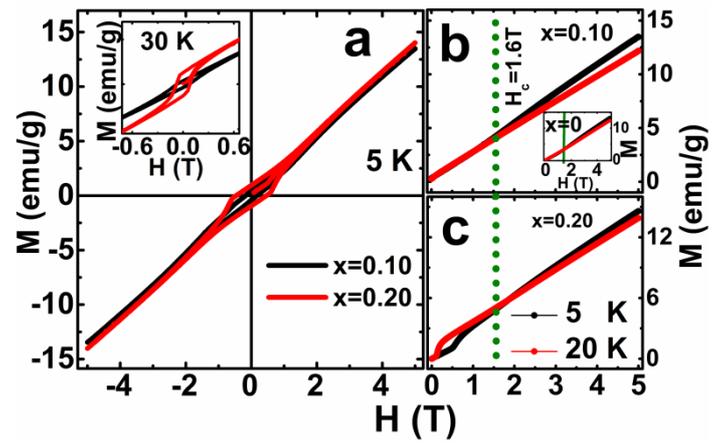

**Figure. 3**

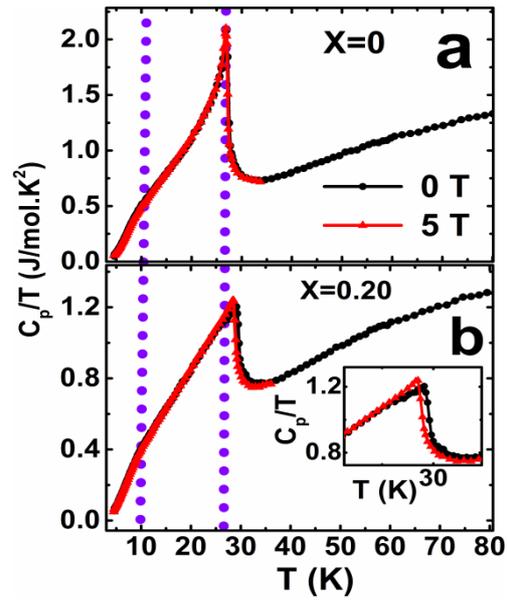

**Figure. 4**

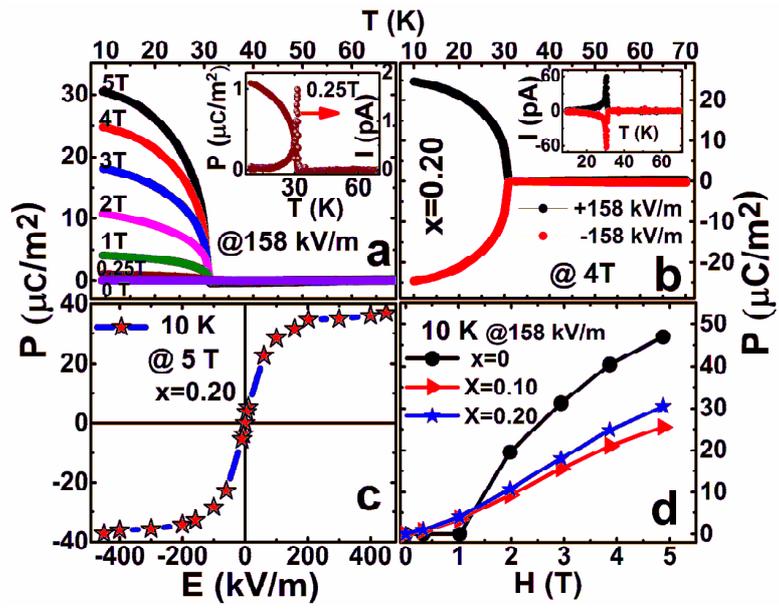

**Figure. 5**